%% file: sapscs.tex
\documentclass{tlp}
\usepackage{latexsym}
\usepackage{alltt}
\usepackage[utf8]{inputenc}

\input{macros}

\title{A structured alternative to Prolog with simple compositional semantics}

\author[António Porto]{ANTÓNIO PORTO\\
LIACC / Department of Computer Science, Faculty of Sciences, University of Porto, Portugal\\
\email{ap@dcc.fc.up.pt}}

\begin{document}

\maketitle

\begin{abstract}
  
  Prolog's very useful expressive power is not captured by traditional logic
  programming semantics, due mainly to the cut and  goal and
  clause order. Several alternative semantics have been put forward,
  exposing operational details of the computation state. We propose
  instead to redesign Prolog around structured alternatives to the cut and
  clauses, keeping the expressive power and computation model but with a
  compositional denotational semantics over much simpler states---just variable
  bindings. This considerably eases reasoning about programs, by programmers
  and tools such as a partial evaluator, with safe unfolding of calls through
  predicate definitions.

  An \textsf{if-then-else}{} across clauses replaces most uses of the cut, but
  the cut's full power is achieved by an \textsf{until} construct. Disjunction,
  conjunction and \textsf{until}, along with unification, are the primitive
  goal types with a compositional semantics yielding
  sequences of variable-binding solutions. This extends to programs
  via the usual technique of a least fixpoint construction. A simple interpreter
  for Prolog in the alternative language, and a definition of
  \textsf{until} in Prolog, establish the identical expressive power of the two
  languages. Many useful control constructs are derivable from the
  primitives, and the semantic framework illuminates the discussion of
  alternative ones.

  The formalisation rests on a term language with
  variable abstraction as in the $\lambda$-calculus. A clause is an abstraction
  on the call arguments, a continuation, and the local
  variables. It can be inclusive or exclusive, expressing a
  local case bound to a continuation by either a disjunction or an
  \textsf{if-then-else}. Clauses are open definitions, composed (and closed)
  with simple functional application ($\beta$-reduction). This paves the way for
  a simple account of flexible module composition mechanisms.

  \cube, a concrete language with the exposed principles, has been implemented on
  top of a Prolog engine and successfully used to build large real-world
  applications.

\vspace{5pt}


\end{abstract}

\section{Introduction}

Practitioners of logic programming have always relied on Prolog's \emph{cut} for
expressive power, not just efficiency, although it destroys the logical reading
of programs and, even worse, is not amenable to a simple compositional reasoning.
The ordering of goals and clauses is also crucial in practice, not just for the
meaning of cuts but to get desired \emph{sequences} of solutions, another
departure from the ideal set-theoretic semantics. Real programs also often rely
on solutions being partial (rather than ground) instances of a call. This aspect
has been given extensive formal treatment in the $s$-semantics approach
\cite{bo_ga_:94:ssata}, and solution order has been captured in an algebra over
solution streams \cite{se_sp_ho:99:alp}, but proposed formalisations of the cut
have serious problems for effective applicability. Most have the inherent
complexity of resorting to a detailed operational state, e.g.\ the decorated SLD
tree \cite{de_mi:88:dosp,sp:00:ogidspc}, success and failure continuations
\cite{dBr_dVi:89:cspc} or a $\pi$-calculus version of such
\cite{li:94:pcsp}. Others are simpler but at the cost of generality by
restricting the use of cut, e.g.\ \cite{And:03:wpspc} where the ``firm'' cut has
just the power of (a limited form of) \mite{}.

In contrast to previous attempts at tackling these problems, our novel approach
avoids a direct characterisation of cut's behaviour and offers instead a few 
linguistic alternatives, which provide Prolog's programming flavour but are
compositional in a simple semantic domain based on variable binding states. One
such alternative is \textsf{if-then-else}, whose semantics
underpins most uses of the cut. Available in Prolog
only within a clause body, we can use it across several clauses,
via a simple reformulation of clause syntax yielding proper compositional
abstractions. But \textsf{if-then-else} does not capture the full power of
Prolog's cut, needed for more generic pruning of solutions. For this we
propose an \textsf{until} construct, that together with conjunction, disjunction
and unification provide the primitive ingredients of the formal machinery. With
these we can define other control constructs such as \textsf{if-then-else},
\textsf{not} or \textsf{var}, and write a very compact interpreter for Prolog
programs, thus showing, since \textsf{until} is easily expressed in Prolog, the
equivalent expressive power of the two languages.

Our linguistic proposal is seemingly simple, both syntactically and (especially)
semantically. We regard it in relation to the cut as structured programming
historically stood to \textsf{goto}, making us firmly believe in its
methodological impact on good programming practice.  The exhibited
compositionality allows programmers to reason locally, i.e.\ disregarding a
goal's context in the code, in terms of \emph{sequences} of variable binding
solutions for the goal, from an initial variable binding setting. Substantially
eased is the task of writing important tools such as debuggers, abstract
interpreters or partial evaluators, as the unfolding of a call through its
predicate definition is sound, being context-independent, in contrast to Prolog.

Our design principles led us to implement a concrete language, called \cube, on
top of a Prolog engine and successfully use it, for a number of years and by many
programmers, to build complex real-world applications such as the online academic
management system at our previous Faculty \cite{po:03:iispp}.

The formalism in the paper is relatively straightforward. We consider a term syntax
with variable scoping (abstraction), for a rigorous account of clause composition and
dynamic variable creation. A (goal) term denotes a \emph{behaviour} function
from an initial \emph{setting}---a variable scope and a substitution---into a
corresponding \emph{outcome} capturing the stream of alternative solutions---a
sequence of settings ending (if finite) with a termination status of finite
failure or divergence. This is recognisable as an abductive semantics for
valuations of goals' variables, a natural way for programmers to understand their
code.  Disjunction, conjunction and \textsf{until} correspond respectively to
notions of sum, product and pruning of outcomes. The semantics of programs with
procedure definitions and calls is given (as usual) as the least
fixpoint of a suitable continuous call step transformer of program
interpretations.

\section{The cut, \textsf{if-then-else} and clause syntax}

Consider this Prolog code for a predicate to \verb+d+elete
\verb+r+epeated \verb+e+lements in a list (for
  variable-tail lists we prefer the notation \texttt{X.L} to
  \texttt{[X|L]}):

\vspace{-3.7pt}
\begin{verbatim}
dre( X.L,   D ) :- X in L, !, dre( L, D ).
dre( X.L, X.D ) :- dre( L, D ).
dre( [],   [] ).
\end{verbatim}
\vspace{-3.7pt}

\noindent
This is a fairly typical case of a definition whose understanding,
although simple, relies crucially on the order of clauses and the
effect of the cut. Even a novice
Prolog programmer will recognise here the implicit pattern of an
\textsf{if-then-else},  with the \textsf{if-then} part stated in the
first clause and \textsf{else} in the next one(s). Indeed, one
can replace the first two clauses with a single one having an
\textsf{if-then-else} body:

\vspace{-3.7pt}
\begin{verbatim}
dre( X.L, Y ) :- X in L -> dre( L, Y ) ; Y=(X.D), dre( L, D ).
\end{verbatim}
\vspace{-3.7pt}

\noindent Although another alternative with cut and
disjunction would work just as well,

\vspace{-3.7pt}
\begin{verbatim}
dre( X.L, Y ) :- X in L, !, dre( L, Y ) ; Y=(X.D), dre( L, D ).
\end{verbatim}
\vspace{-3.7pt}

\noindent we must realise that the \textsf{if-then-else} body can soundly replace
appropriate calls for \texttt{dre} when doing partial evaluation of a program,
whereas the one with the cut cannot, as the cut would apply to a different
definition scope. One should, therefore, definitely prefer a structural
\textsf{if-then-else} to its unstructured implementation with cut. For all but
very simple definitions, however, it is inconvenient to trade multiple clauses
for one single clause body, following the general principle of keeping local definitions
concise in order to ease reasoning about program behaviour. What is needed, then,
is a more direct structural syntax for clauses that yields an \mite{} meaning,
giving us the expressive power to cascade \mite{}s across clauses. Here is our
concrete alternative syntax for the original first clause of \texttt{dre}:

\vspace{-3.7pt}
\begin{verbatim}
dre( X.L, D )  <-  X in L  <>  dre( L, D ).
\end{verbatim}
\vspace{-3.7pt}

We call this an \emph{iff-clause} due to the analogy with a biconditional
(under an implication), in this case $(\texttt{X in
  L})\imp(\texttt{dre(X.L,D)}\bic\texttt{dre(L,D)})$.  Procedurally, if a call
unifies with the head \verb+dre(X.L,D)+ and the (\verb+X in L+) condition
succeeds, then solving \verb+dre(L,D)+ is the only way to solve the call;
otherwise, the call's solutions must come from the next clauses. Such a clause
therefore represents an \textsf{if-then-else} statement abstracted on the
\textsf{else} part, to be plugged with the statement corresponding to the
continuation of the definition. This compositionality for clauses is formalised
in the paper by considering an extended syntax incorporating variable abstraction
and application as in the $\lambda$-calculus. To make the abstract nature of
clauses even more apparent we introduce syntax to lump them under a single occurrence
of the procedure name, as follows (sugared syntax on the left, unsugared 
on the right).

\vspace{-3.7pt}
\begin{verbatim}
dre                             dre
::  X.L,   D  <-  X in L        ::  X.L,   D  <-  X in L
              <>  dre( L, D )                 <>  dre( L, D )
                                ..  X.L, X.D  <-  true
..  X.L, X.D  <>  dre( L, D )                 <>  dre( L, D )
                                ..  [],  []   <-  true
..  [],   []  !.                              <>  true
\end{verbatim}
\vspace{-3.7pt}

The programming style promoted by using clauses is to split the definition of a
procedure into \emph{cases}, typically related to certain patterns of
arguments. Given the semantic possibility of multiple solutions, the central
decision to be made when thinking about a case is whether it should be
\emph{exclusive}, i.e.\ precluding subsequent cases from being considered, or
\emph{inclusive}, when alternative solutions may come from subsequent cases. In
\cube{} this is expressed by choosing one of two clause formats, an
\emph{iff-clause} \texttt{A<-C<>B} or an \emph{if-clause} \texttt{A<-B} (akin to
\texttt{A:-B} in Prolog), standing respectively for an implicit \mite{} or a
disjunction, with the \textsf{else} or second disjunct a placeholder for
the rest of the procedure definition. The choice is exemplified by comparing a
multi-solution procedure for producing members of a list with one that just
checks for membership of a given item (as in Prolog, $H$ stands for
$H$\texttt{<-true}).

\vspace{-3.7pt}
\begin{verbatim}
member                            has_member
::  X._, X                        ::  X._, X  !
..  _.L, X  <>  member( L,X ).    ..  _.L, X  <>  has_member( L,X ).
\end{verbatim}
\vspace{-3.7pt} The syntax promotes quick recognition of whether a clause is
exclusive or inclusive, through the presence or absence of a \emph{single}
occurrence of \texttt{<>} (or its
sugared variant
\texttt{!}). The syntax remains very close to that of Prolog; the whole point is
to provide a much cleaner semantics, as the paper will show, with minor syntactic
adjustments.

\section{The true power of cut: \textsf{until}}
\label{sec:until}

The power of \textsf{if-then-else}, and therefore of iff-clauses, is enough to define many other useful
control constructs found in Prolog, such as \texttt{once}, \texttt{not}
or \texttt{var}. If, however, we want to define operations that can stop the production of
solutions after possibly more than one, \textsf{if-then-else} is no
longer enough and we need a more powerful \textsf{until} construct. A
simple and instructive way to convey its meaning is to write its
definition in Prolog:

\vspace{-3.7pt}
\begin{verbatim}
Solve until Stop :- Solve, ( Stop, ! ; true ).
\end{verbatim}
\vspace{-3.7pt}
The solutions for (\texttt{Solve until Stop}) are the initial ones of \verb+Solve+ for
which \verb+Stop+ fails, plus
the first (if ever) for which \verb+Stop+ succeeds, and then no more. Notice
that we need to use a cut in a conjunctive disjunct of a disjunctive conjunct,
precisely the kind of context where a cut achieves its full power.

Having \texttt{until} as a basic primitive, along with conjunction,
disjunction and unification, we can implement \textsf{if-then-else}. We take this
opportunity to propose the notation (\texttt{If -> Then -; Else}), avoiding
Prolog's bad overload of ``\verb+;+'' for both disjunction and the
\textsf{else} separator. We use \texttt{until} as an infix operator binding
tighter than the comma, i.e.\ \texttt{(A until C, B)} $\equiv$ \texttt{((A until
  C),B)}. The implementation uses an auxiliary variable \texttt{R} to
convey the result of the test (\texttt{t} for taking the
`\textsf{then}' branch, \texttt{e} (\textsf{else}) otherwise), bindable only once
$( \texttt{X until true}\equiv\texttt{once X})$.

\vspace{-3.7pt}
\begin{verbatim}
( If -> Then -; Else ) <> ( If, R=t ; R=e ) until true,
                          ( R=t, Then ; R=e, Else ).
\end{verbatim}
\vspace{-3.7pt}

Another useful derived construct is \textsf{unless}, akin to
\textsf{until} but failing rather than succeeding after the stop
condition holds. We implement it with a result variable
that is bound only upon a successful test, then causing final failure:

\vspace{-3.7pt}
\begin{verbatim}
Solve unless Stop  <>  Solve until ( Stop, R=f ), R=s.
\end{verbatim}
\vspace{-3.7pt}
With \texttt{unless} we can write e.g.\ a clean local read-process
repeat-fail loop:

\vspace{-3.7pt}
\begin{verbatim}
( repeat, read(Item) ) unless Item=end_of_file, process(Item), fail
\end{verbatim}
\vspace{-3.7pt}

To show that \textsf{until} really holds the full power of Prolog's cut we write
an interpreter for Prolog in our cut-free language. The presentation is
simplified by the use of iff-clauses and \textsf{unless}, knowing that both are
implementable with \textsf{until}. We assume a unary procedure \verb+system+ for
identifying system calls (\verb+true+ being one).

\vspace{-3.7pt}
\begin{verbatim}
execute
::  G  <>  exec(G,R) unless R=fail.
exec
::  (A,B), R  <>  exec(A,RA), ( RA=fail, R=fail ; exec(B,R) )
..  (A;B), R  <>  exec(A,R) ; exec(B,R)
..  (!),   R  <>  R=succ ; R=fail
..  G,  succ  <-  system(G)
              <>  G
..  G,     R  <>  ( clause(G,B), exec(B,R) ) unless R=fail.
\end{verbatim}
\vspace{-3.7pt}
\noindent The main idea is that \texttt{exec} may succeed with two results in its
second argument: \texttt{succ} signals true success, coming from the base
\texttt{true} (a \texttt{system} call) or a cut's initial success; backtracking
past a cut, though, actually succeeds again but with the \texttt{fail} result;
this fake success is propagated through conjunctions (from or bypassing the
second conjunct) and disjunctions (any branch), eventually exiting the
\texttt{exec(B,R)} call of a \texttt{clause} body \texttt{B}, at which point
the \texttt{unless} condition succeeds for the first time, immediately calling
off the production of more solutions for \texttt{G} from any pending choices in
\texttt{exec} or \texttt{clause} (similarly for the top goal in the
\texttt{execute} clause).

We now turn to the problem of formally characterising the
compositional semantics behind this reconstruction of Prolog along
structured principles, aiming at a precise and clear understanding of
program behaviour.

\section{Syntax}

It helps in the formalisation to consider an \emph{abstract} syntax embodying the
structural principles onto which the concrete syntax maps. We adopt a simple
abstract syntax that is universal, given its suitability to encode the syntax of
various formal calculi in the absence of predefined semantic roles for its
constructs.

\subsection{Terms}

  We use variables and a special null as basic terms, and three
constructors: for pairing two terms (which, together with the null, provide
lists), for applying a constant to a list (giving us rooted terms) and for
abstracting a variable in a term (scoping).

\begin{definition}
  The identifiers are a set of \emph{constants} $\C$ and a
  totally ordered countably infinite set $\V$ of \emph{variables} disjoint from
  $\C$.  The \emph{terms} are the smallest set $\T$
  satisfying\\[-13pt]
  \begin{eqnarray}
\T = \{\nil\}\cup \V\cup (\T\times \T)
\cup({\C}\times{\LL}) \cup(\{\lambda\}\times{\V}\times{\T})    
  \end{eqnarray}\\[-13pt]
where $\nil$ is the
\emph{null}, $(t,t')\in(\T\times\T)$ is a \emph{pair}, $(\appl
cl)\in({\C}\times{\LL})$ is an \emph{application}, $(\abst
vt)\in(\{\lambda\}\times{\V}\times{\T})$ is an \emph{abstraction}, and the
\emph{lists} are the smallest set $\LL\subset\T$
satisfying $\LL=\{\nil\}\cup(\T\times\LL)\,$.
\end{definition}

Notice that formally a constant $c\in\C$ is not a term, but $c$ in the
concrete syntax corresponds to the term $\appl c\nil$. The familiar
notation $c(t_1,\ldots,t_n)$ for structured
terms in the concrete syntax corresponds abstractly to $\appl
c{(t_1,(\ldots,(t_n,\nil)\ldots))}$, also represented in
our metalanguage by $\appl c{[t_1,\ldots,t_n]}$, following Prolog's
tradition. This inevitably reminds us of the \textsf{univ} ({\tt=..})
system predicate in Prolog, which does indeed
correspond to the construction/deconstruction of an application term
from/into its two components.

Abstraction terms have the usual intrinsic syntactic property of scoped variable
capture, seen in the next definition.

\begin{definition}
\label{def:free}
  The \emph{free variables} $\fv t$ of a term $t\in\T$ are defined
  recursively as follows.
  \begin{xalignat*}{2}
\fv{\appl cl}&=\fv{l}	&\qquad \fv{\nil}&=\emptyset\\
\fv{a,b}&=\fv a\cup\fv b &\qquad \fv{v}&=\{v\}\qquad\mathrm{if}\quad v\in\V\\
\fv{\abst vt}&=\fv{t}\,\setminus\{v\} &&
  \end{xalignat*}
\end{definition}

In this universal syntax we are able to encode type-free $\lambda$-calculus terms
using \ $\encode{\abst ve}=\abst v{\encode e}$ and $\encode{(a)b}=(\encode
a,\encode b)$, predicate calculus formulae with \ $\encode{\forall
  vF}=\appl\forall[{\abst v{\encode{F}}}]$ (assuming $\forall\in\C$), etc., with
no predefined semantics for the syntax on its own. Only for a certain intended
use---in context---may terms and constants acquire a particular meaning,
formalised in a semantics.

In the sequel we shall use both abstract and concrete syntax, according
to contextual convenience.

\subsection{Clauses, procedures and programs}
\label{sec:clauses}

The true power of any programming paradigm comes from the ability to
define procedures and interpret certain expressions as procedure calls, so
we proceed with a syntactic characterisation of procedures in our
framework.

A procedure definition is built from sequences of clauses (in the concrete
syntax) and captured in a single term (in the abstract syntax) that provides
the semantics of calls to the procedure. We use certain terms to encode
open (partial) definitions, such as clauses, and a simple syntactical composition for
building larger open definitions from smaller ones. Closing an open definition is
a simple operation.

The main intuition is that a procedure call is an \emph{application} term
with a root constant and a list of arguments, whereas a closed procedure
definition for the root constant is an \emph{abstraction} term
 to be applied (in the
$\lambda$-calculus $\be$-reduction sense) to the call's argument list, resulting in
a term to be further evaluated.

For example, the following definition, taken as closed,

\vspace{-3.7pt}
\begin{verbatim}
int  ::  0
     ..  s(X)  <-  int( X ).
\end{verbatim}
\vspace{-3.7pt} associates \verb-int- to the abstraction term $\abstt{A}{( A=[0];
  $\abstt{X}{( A=[s(X)], int(X) )}$ )}$. The term is very reminiscent of Clark's
completion \cite{cl:78:nf}, as should be expected in this example. But whereas
Clark's construction always uses disjunction when composing clauses, ours may use
\textsf{if-then-else} rather than disjunction, when composing exclusive rather
than inclusive clauses. Let us look, then, at the process of building a complete
procedure definition.

A clause stands for (a base case of) a \emph{partial} definition---an abstraction
term abstracting away the argument list and the alternative continuation of the
definition, plus any local variables, under which we find either a disjunction or
an \textsf{if-then-else}, respectively for if-clauses or iff-clauses, as this
general translation shows:\\[-7pt]

\hspace{-4.5mm}\begin{tabular}{ll}
$a_1,\ldots,a_n$ \texttt{<-} $B$ &
\ \ $\abst{\texttt{A}}{\abstt{D}{( $\lambda  v_1\cdots\abstt{$v_k$}{(
        A=[$a_1,\ldots,a_n$],$B$ ) ; D )}$}}$ \\
$a_1,\ldots,a_n$ \texttt{<-} $C$ \texttt{<>} B &
\ \  $\abst{\texttt{A}}{\abstt{D}{( $\lambda  v_1\cdots\abstt{$v_k$}{(
        A=[$a_1,\ldots,a_n$],$C$ -> $B$ -; D )} $)}}$
\end{tabular}\\[-2pt]

\noindent with $v_1,\ldots,v_k$ the variables in each clause.
Take as an example the following two clauses for a binary procedure:
\vspace{-3.7pt}
\begin{verbatim}
X, a  <-  r(X).
1, b  <>  true.
\end{verbatim}
\vspace{-3.7pt}
The if-clause is
equivalent to the following partial definition\\[-13pt]
$$p_1=\abstt{A}{$\abstt{D}{( $\abstt{X}{( A=[X,a], r(X) )}$; D )}$}\,,$$
while the iff-clause represents this other one\\[-13pt]
$$p_2=\abstt{A}{$\abstt{D}{( A=[1,b] -> true -; D )}$}\,.$$

The effect of putting one clause after another can be defined as a
generic syntactic composition of partial definitions,\\[-13pt]
\[p'\maft
p=\abstt{A}{$\abstt{D}{( $\apply{(\apply{p}\texttt{A})}{(\apply{(\apply{p'}\texttt{A})}\texttt{D})}$ )}$}\]
where $\apply{(\abst xt)}a$ stands for
the replacement by $a$ of all free occurrences of $x$ in $t$, i.e.\
the equivalent of applying the $\beta$-rule in the $\lambda$-calculus to
$(\lambda x.t)a$.
 
In our example we can compose in two ways:\\[-13pt]
\begin{eqnarray*}p_2\maft p_1&=&\abstt{A}{$\abstt{D}{( $\abstt{X}{(A=[X,a], r(X))}$; (A=[1,b] -> true -; D) )}$}\,,\\
p_1\maft p_2&=&\abstt{A}{$\abstt{D}{( A=[1,b] -> true -;
($\abstt{X}{(A=[X,a], r(X))}$; D) )}$}\,.
\end{eqnarray*}

Typically clauses are composed in their textual order. In a modular
version of the language one may wish to compose a generic definition after
a more specific one (defaults after overriding exceptions) or the other way
around (specific cases uncovered by general ones).

Closing a partial definition is simple. With
$\denot{\texttt{fail}}(s)=\ff$ (e.g.\ $\texttt{fail}\equiv\texttt{a=b}$),\\[-13pt]
\[\mclo p=\abstt{A}{( $\apply{(\apply p{\texttt{A}})}{\texttt{fail}}$ )}\,.\]

\noindent After closing a definition, the abstraction of the definition's
continuation has vanished. There is one outer abstraction on the argument list,
and any remaining abstractions are for local clause variables.

A \emph{program} is formally a mapping from constants to appropriate abstraction
terms standing for closed procedure definitions.  Cases for
different argument arities under the same constant are possible, as in
Prolog; grouping them together, rather than by
constant/arity pairs, is just a technical convenience.

\section{Semantics}

We are interested in defining \emph{denotational} semantics for our terms,
capturing their solution-producing behaviour when invoked as goals (\emph{tasks},
we prefer to call them).

What are the basic intuitions for defining the semantics? A task generally has
variables, and its behaviour results in solutions for them, expressed
as constraints on the variables' possible valuations (as ground terms), a notion we
call a \emph{setting}. A task is launched in the context of an initial setting
(a previously executed task may share some of the variables) and
any solution is a  possibly more constrained setting satisfying the initial one. A
task may produce more than one solution, in a well-defined order. The sequence of
solutions may be infinite, or else the task ends up by either finitely failing or
diverging in the search for more solutions.

\subsection{Settings, outcomes and behaviours}

For constructing our semantic domain, then, we wish to define a
setting providing partial information on some variables' possible
valuations. It should satisfy the very abstract characterisation we gave in
\cite{mo_po:98:ebac} of a structure with a partial order of entailment and
consistency completeness, adequate for arbitrary constraint logic
programming languages. In this paper we restrict ourselves to a Prolog-like
language handling only identity constraints through unification, and define
settings accordingly.

Formalising a setting of identity constraints may vary in the degree of
abstraction. In the WAM \cite{ai-ka:91:wamtr} implementation of Prolog we can see
settings as accumulated equations of variables to terms, but these are too
concrete. Striving to approach full abstraction, we opt instead for substitutions
expressing the solved form of those equalities, actually the view commonly held
by programmers. However, our settings will formally differ from standard
substitutions, to cater for two needs (clarified ahead): ideal infinite terms, and an
explicit scope of variables.

\begin{definition}
  The \emph{ideal terms} $^I\T$ are the largest (not smallest) solution set for
  $\T$ in equation (1). Definition \ref{def:free} of free variables carries
  over from $\T$ to $^I\T$ by assuming the smallest set satisfying the equations.
  The \emph{substitutions} $\Ss$ are the mappings $\s:V\imp{}^IT$ from a finite
  set of variables $V\fss\V$ to ideal terms where they do not occur free,
  $v'\in\fv{\s(v)}\Imp v'\not\in\dom\s$, and not mapping a variable to a smaller
  one (they are totally ordered),  $\s(v)\in\V\sImp
  v>\s(v)$. Considering $\fv\s=\set{\fv{\s(v)}}{v\in\dom\s}$ the free variables
  of a substitution $\s\in\Ss$, the \emph{settings} $\S$ are the pairs $\stng
  V\s$ of a variable \emph{scope} $V\fss\V$ and a substitution $\s\in\Ss$ under
  the scope, with $\dom\s\cup\fv\s\subseteq V$.
\end{definition}

\noindent The difference $^I\T\setminus\T$ between ideal and regular terms are
the so-called \emph{infinite} terms. Their possible appearance in substitutions
reflects most Prolog implementations, that by omitting the expensive occurs
check in unification may generate solutions corresponding to infinite
\emph{rational trees}, as indeed proposed by Prolog's inventor
\cite{co:82:pif}. The definition of substitution conveys the
idea that the implicit equations are sufficiently unfolded into a ``solved
form''. For example, $\s=\{\soleq X{f(Y)},\soleq Ya\}$ is not a substitution, as
the domain variable
{\tt Y} occurs in $\s({\tt X})$. The requirement on variable order aids in the
determinacy of settings, by following the WAM's policy when binding a pair of
free variables.

Now we come to the notion of \emph{outcome}, to express the deterministic result
of executing a task in the context of a given setting, as a sequence of solutions
and termination status. For example, the task \verb+member(1.X,Y)+ launched in
the setting $\stng{\{{\tt X},{\tt Y}\}}{\{\soleq X{[2]}\}}$ yields a first
solution $\stng{\{{\tt X},{\tt Y}\}}{\{\soleq X{[2]},\soleq{Y}{1}\}}$, then after
backtracking a second one $\stng{\{{\tt X},{\tt
    Y}\}}{\{\soleq{X}{[2]},\soleq{Y}{2}\}}$, and if retried again finitely
fails. A semantics based on sequences of solutions is not new, having been
proposed for algebraic logic programming \cite{se_sp_ho:99:alp}; ours differs in
the form of those solutions (settings) and the inclusion of termination
status.

\begin{definition}
  The \emph{outcomes} are $\Out=\Out^f\cup\Out^s$, with $\Out^f=\{\ff,\infty\}$
  the \emph{final} outcomes and $\Out^s$ the \emph{successful} outcomes---the greatest set satisfying $\Out^s=\set{\sout
    so}{s\in\Stng,\,o\in\Out}$.
\end{definition}

\noindent The symbols $\ff$ and $\infty$ represent the final outcomes of,
respectively, finite failure and divergence. A successful
outcome is a non-empty sequence of solutions, either
infinite or terminated by a final outcome.

We want to capture the variability of settings in which a task is executed,
affecting its outcome. It becomes relevant to define the entailment relation on
settings, to support the intuition that the outcome of a task
can only have solutions with equal or stronger constraints than the
setting at the start, i.e.\ entailing it.

\begin{definition}
  For any ideal term $t$ and substitution $\s$, let $t[\s]$ denote the ideal term
  obtained by replacing in $t$ any occurrence of a variable $v\in\dom\s\cap\fv t\,$ by $\s(v)$.
 The \emph{entailment} $\s'\vdash \s$
  between substitutions is defined by $\s(v)[\s']=\s'(v)$ for every
  $v\in\dom\s\subseteq\dom{\s'}$. For settings, $\stng{V'}{\s'}\vdash\stng{V}{\s}$ whenever
  $V'\supseteq V$ and $\s'\vdash\s$.
  We say that $o\in\Out$ is an outcome \emph{upon} a setting
  $s\in\Stng$, written $\ous os$, if all the solutions entail it, i.e. with $S(o)$
  denoting the \emph{set} of settings in $o$ we have $s'\in
  S(o)\Imp s'\vdash s$.
\end{definition}

Just before finally defining behaviours for tasks, we remark that the setting in which a task
is executed must have a scope covering the task's free variables, but
possibly also some other variables from the task's original lexical scope (typically a
program clause). It becomes convenient to index behaviours by sets of
variables covered by, rather than equal to, the scopes of the involved settings.

\begin{definition}
  $\S_V=\set{\stng{V'}\s\in\S}{V\subseteq V'\fsubset\V}$ are the settings
  \emph{covering} $V$. The \emph{behaviours} are $\B=\bigcup_{V\fsubset\V}\B_V$, with
  $\B_V$, the behaviours \emph{for} $V$, being the mappings $\be:\S_V\imp\Out$ from
  settings covering $V$ to outcomes upon them, i.e.\ such that $\ous{\be(s)}s$.

\end{definition}

The denotational semantics for terms, taken as tasks, is a mapping $\denot
.:\T\imp\B$ into behaviours for the terms' free variables, i.e.\ $\denot
t\in\B_{\fv t}\,$. Notice that, according to the given definitions, the denotation
of a task yields outcomes for initial settings whose scope is a superset of the
task's free variables. The denotation mapping $\denot .$ must satisfy certain
equations for a class of \emph{special} terms $\T^\ast$ that have a predefined
compositional way of being interpreted as tasks, whereas the other terms are
interpreted in the context of a given program. We now proceed to introduce the
members of $\T^\ast$, along with their fixed denotation equations. To lighten the
presentation we use concrete infix operator syntax (rather than abstract) for
such terms.

\subsection{Disjunction}

The denotational semantics of a disjunctive term  $(a\obor b)\in\T^\ast$ is given through a
semantic \emph{sum} operation $\ssum:\Out\times\Out\map\Out$, as follows.

\vspace{-15pt}
\begin{eqnarray*}
\denot{a\obor b}(s)=\denot{a}(s)\ssum\denot{b}(s) \qquad\qquad \infty\ssum o'&=&\infty\\
\ff\ssum o'&=&o'\\
  (\sout so)\ssum o'&=&\sout s{(o\ssum o')}
\end{eqnarray*}
Each disjunctive sub-term is evaluated in the same initial setting---the essence
of the backtracking process that implements this semantics. Failure of the first disjunct
leads to collecting the solutions of the second, and divergence of the first
naturally extends to the whole disjunction.

\subsection{Conjunction}

For the semantics of a conjunctive term $(a\oband b)\in\T^\ast$ we use a semantic
\emph{product} operation $\sprod:\Out\times\B\map\Out$ that relies on the sum, as
follows.

\vspace{-15pt}
\begin{eqnarray*}
\denot{a\oband b}(s)=\denot{a}(s)\sprod\denot{b} \qquad\qquad
\infty\sprod \be&=&\infty\\
\ff\sprod\be&=&\ff\\
  (\sout so)\sprod \be&=&\be(s)\ssum(o\sprod\be)
\end{eqnarray*}
The evaluation of the second conjunct is performed upon each solution
of the first (yielding stronger solutions). As expected, both failure and
divergence of the first are absorbing.

\subsection{Until}

The intended behaviour of a term $(a\obunt b)\in\T^\ast$ is to provide the
solutions of $a$ but checking, upon each one, whether $b$ has a successful outcome,
in which case the corresponding solution is the last to be provided. This is
achieved by a \emph{pruning} operation $\sprun:\Out\times\B\map\Out$, as follows.

\vspace{-15pt}
\begin{eqnarray*}
\denot{a\obunt b}(s)=\denot{a}(s)\sprun\denot{b} \qquad\;\;
\infty\sprun \be&=&\infty\\
\ff\sprun\be&=&\ff\\
  (\sout so)\sprun \be&=&\caseif{
     \infty\cond{\be(s)=\infty}\\
      \sout s{(o\sprun\be)}\cond{\be(s)=\ff}\\
      \sout{s'\!}\ff\cond{\be(s)=\sout{s'\!}{o'}}}
\end{eqnarray*}
Notice how the first successful solution $s'$ of the pruning condition is
taken as the final global solution (it entails the solution $s$ for
the pruned task), discarding further solutions from both the condition
($o'$) and the pruned task ($o$). As expected, failure of the
pruned task and divergence of either task are absorbing.

\subsection{Unless  and if-then-else (revisited)}
\label{sec:revis}

Although from the perspective of minimal semantic ingredients 
\texttt{unless}  and  \mite{} are not primitive constructs, being definable through
\texttt{until}, it is enlightening to see them defined directly in our semantic
framework.

For \texttt{unless} we need a very simple variation on the pruning operator
of \texttt{until}, obtained by replacing, in the last line of the definition
above, $\sout{s'\!}\ff$ with just $\ff$.

The implementation of \mite{} given in section \ref{sec:until} matches this definition:\\[-13pt]
\[
\denot{\mathit{if}\,\mtt{->}\,\mathit{then}\,\mtt{-;}\,\mathit{else}}(s)=\caseif{
\infty\cond \denot{\mathit{if}}(s)=\infty\\
\denot{\mathit{else}}(s)\cond{\denot{\mathit{if}}(s)=\ff}\\
\denot{\mathit{then}}(s')\cond{\denot{\mathit{if}}(s)=\sout{s'\!}o}}
\]
Ahead in section \ref{sec:alter} we discuss an alternative meaning adopted by
other languages.

\subsection{Unification - the prime mover}

Any computational engine using the given compositional
interpretation for the three operators must also define the
denotational semantics for some terms that act as the basis for
change, building stronger settings from
previous ones.
Here we assume as basic just the operation of
\emph{unification} of two terms, captured syntactically by special
terms $(a\obeq b)\in\T^\ast$.\\[-13pt]
\[
\denot{a\obeq b}(\stng V\s)=\caseif{
\infty\cond \Unf[\not]ab{}\\
\ff\cond{\Unf ab\bot}\\
\sout{(\stng V{\s'})}{\ff}\cond{\Unf ab{\s'\in\Ss}}}
\]

The unification $U(a,b,\s)$ of $a$ and $b$ under $\s$ may succeed with an equal
or stronger substitution $\s'\vdash\s$, yield failure ($\bot$) or diverge (when
unifying infinite terms).  A successful unification $\Unf
ab{\s'}$ yields the \emph{least} substitution $\s'$, in the partial order of
entailment, that makes $a$ and $b$ identical, $a[\s']$ =
$b[\s']$. We can formalise unification by an inductive definition, adapting to
our more ideal framework the classical definition introduced by Robinson for the
predicate calculus but without the occurs-check, in the spirit of Colmerauer's
suggestion and in accordance with most Prolog implementations, although not
mandated by its standard \cite{de_ed-db_ce:91:ps}. This being quite well known we
omit the details.

Although abstraction terms are included in our syntax
and implicit in clauses, they never actually appear inside
unification, if no explicit concrete syntax for them is available. Otherwise
unification has to handle also $\alpha$-conversion equivalence.

\subsection{Atomic terms}

In our abstract syntax the only atomic terms are the variables
$v\in\V$ and the null $\nil$. Since pairs were given the
semantics of conjunction, the natural extension to lists is
to treat the null as special ($\nil\in\T^\ast$), with the idle successful outcome:\\[-13pt]
\[\denot{\nil}(s)=\sout s\ff\]
We equate \texttt{true} in the concrete syntax to the abstract null $\nil$.

Variables are also special ($\V\subset\T^\ast$), being interpreted under the
setting. A resulting free variable has no procedural meaning, yielding a finite failure
outcome.
$$v\in\V\quad\Imp\quad\denot v(\stng V\s)=\caseof{
\ff \cond{\s(v)\in\V}\\
\denot{\s(v)}(\stng V\s) \other}$$

\noindent This simple definition captures the quite useful higher-order
feature of Prolog-like languages exemplified by
\verb=( build_task(Data,Task), Task, process(Data) )=, where the \verb+Task+
variable is first
bound to a term (sharing variables with a \verb+Data+ pattern) that gets to be
executed as a task
(instantiating the \verb+Data+ to be processed).

\subsection{Procedure call}

The denotational semantics of a non-special term $(\appl
pa)\in\T\setminus\T^\ast$, interpreted as a procedure
call, is parametric on a given program $P$, as follows.
$$\denot{\appl pa}_P=\denot{\apply{P(p)}{a}}\;.$$
We have seen that $P(p)$, the closed
definition for $p$ in the program $P$, is
always an abstraction term whose outermost abstraction is on the argument list.
Taking the example of \texttt{int} in section \ref{sec:clauses}, a call
$\texttt{int(s(a))}\equiv\appl{\texttt{int}}{\texttt{[s(a)]}}$ results in
applying the {\tt int}
abstraction to the call's argument list
\texttt{[s(a)]}, resulting in the task\\[-13pt]
\[\texttt{( [s(a)]=[0]; $\abstt{X}{( [s(a)]=[s(X)], int(X) )}$ )}\]
whose behaviour, since that of the first disjunct yields $\ff$ (unification
failure), is the behaviour of the inner abstraction $\abstt{X}{( [s(a)]=[s(X)],
  int(X) )}$.  This term has an implicit existential reading of  \texttt{X} as a clause
variable, and its launch as a task starts by
replacing the abstracted variable in the inner term, before executing it, with a
fresh new variable for the current setting---the analogue of using clause
variants in resolution---as defined next.

The denotational equation given above states the correctness of unfolding a
procedure call with its procedure definition. This is what makes e.g.\ partial
evaluation much easier for this structured language than for Prolog, where cuts
in procedure definitions make such unfolding unsound due to the scope extrusion of
the cuts.

\subsection{Abstraction}

Abstraction terms $(\abst xt)\in\T^\ast$ come from
clause variable scoping in procedure definitions.  Invoked as tasks they give
rise, as mentioned, to
the creation of new variables for the solutions of the clause case. Formally,
$$\denot{\abst vt}(\stng V\s) = \denot{\apply{(\abst vt)}{\next V}}
(\stng{V\cup\{\next V\}\,}\s)$$
with $\next V$ being the function, implicitly defined by the countable order on
$\V$, that returns the least variable greater than those in $V$. This justifies
the need for the scope in settings, formalising how the stack grows in the WAM
implementation.

\subsection{Fixpoint semantics}

If a term's structure is composed solely of special terms, the corresponding
recursive equations uniquely define the term's
denotational semantics. This is no longer the case for procedure calls,
because of the circularity introduced by recursive definitions. The
standard solution in logic programming is to define a mapping from programs to
continuous operators on the possible interpretations and give the semantics
of a program as the least fixpoint of its operator \cite{v-em_ko:76:splpl}. We will
proceed likewise, but for our different semantic domain.

The \emph{interpretations} $\I$ are the functions $I:\T\imp\B$ that map each term
into a behaviour for its free variables, $I(t)\in\B_{\fv t}\,$,
and satisfy the equations given for \emph{special} terms when $I$ is taken
for $\denot .$. Interpretations differ, then, in the behaviours of the
non-special terms, i.e.\ the procedure calls.

We define a partial order on interpretations based on that of
outcomes,$$I\sqsubseteq I'\quad\Bic\quad\forall t\in\T\;\,\forall
s\in\Stng_{\fv{t}}\;\,I(t)(s)\sqsubseteq I'(t)(s)\;,$$
the partial order on outcomes being the greatest relation
that satisfies
$$o\sqsubseteq o'\quad\Bic\quad(o=\infty)\scor(o=o'=\ff)\scor(o=\sout
su,\,o'=\sout s{u'},\,u\sqsubseteq u')\;.$$ 
Notice that having $o\sqsubseteq o'$ with
$o\neq o'$ is possible only when $o$ ends in $\infty$ after a common (finite) prefix with
$o'$. Intuitively this can be understood as $o$ and $o'$ being partial outcomes
for the same task but with fewer computation steps available to produce $o$,
reflected in the divergence ``termination''.

The \emph{call step transformer} $\stp\in\I^\I$ for a given program $P$
maps an interpretation $I$ into an interpretation $\stp(I)$ such that,
for any non-special term $(\appl pa)\in\T\setminus\T^\ast$,
$$\stp(I)(\appl pa)=I(\apply{P(p)}a)\;.$$
$\stp$ is continuous (we omit the proof) and has a least fixpoint which is the
semantics (the model) of the program $P$, satisfying the semantic equation
 for  procedure calls.

\subsection{Abduction}

An interesting semantic insight is to interpret the behaviour of unification
tasks as performing abduction \cite{ka_ko_to:93:alp}. A setting $\stng V\s$ can
be thought of as a theory $\Theta(\stng V\s)=\set{v\mbox{\tt =}t}{(v,t)\in\s}$ in
a variable-free logic language where $V$ are considered Skolem constants
interpreted in the realm of ideal terms, and the single predicate symbol `{\tt=}'
is interpreted under the standard equality axioms $\E$. Whenever $\denot{a\obeq
  b}(s)=\sout{s'\!}\ff$ we can see that $\Theta(s')$ is a minimal consistent
extension of $\Theta(s)$ such that $\Theta(s')\cup\E\models a\mbox{\tt=}b$, and
if no such extension exists then $\denot{a\obeq b}(s)=\ff$. We spot here the
hallmarks of abduction, and indeed the outcome solutions may be seen as the
abductive extensions that make true the equality statements implicit in the
unifications along the way.

The pruning semantics, interestingly, can establish
another way of relating tasks to abduction. Calls to the procedure
(\texttt{possible X <> not not X}), and to (\texttt{var X <> possible X=A,
  possible X=b}), are actually statements about abducibility in the
\emph{current} setting, rather than requirements for abductive extension. The
given \texttt{var} definition reads directly as ``it is currently possible to
abduce equality of \verb=X= to both \verb=a= and \verb=b=''. This semantic
dependency on the current setting, rather than a final solution, clearly explains
why conjunction is not commutative, e.g.\ $\texttt{(var(X),X=a)}$.

\section{Language design: alternatives and extensions}
\label{sec:alter}

We presented \textsf{until} as the basic semantic ingredient for achieving the
power of Prolog's cut, but in practice several derived constructs are available
to the programmer, such as {\tt not}, {\tt once} or \mite. The latter is
pervasive, being the implicit underpinning of exclusive clauses, the vast
majority in real programs. The meaning we took for \mite{} is expressed in its
definition in section \ref{sec:revis}---only the first solution of the
$\mathit{if}$ condition, if it exists, is retained as initial setting for the
$\mathit{then}$ part. Alternatively, the designers of e.g.\ NU-Prolog
\cite{NUProlog} and Mercury \cite{Mercury} have chosen, on the grounds of it
being more declarative and logically sound, to use \emph{all} solutions of the
condition. This is just as easy to define, using
$\denot{\mathit{if}}(s)\sprod\denot{\mathit{then}}$ instead of
$\denot{\mathit{then}}(s')$ in the third case of the definition.  Our choice was
pragmatic, being aligned with Prolog and validated by usefulness in
applications. We never encountered a real need for the supposedly more
declarative reading of \mite, even though consciously on the lookout for it. We
did provide in \cube{} a related $\mathit{otherwise}$ construct yielding all
solutions of its $\mathit{if}$ part, but also found it wanting of applicability.
Interestingly, we point out that while our reading of \mite{} can be implemented
with unification, conjunction, disjunction and \texttt{until}, this is not the
case for the alternative reading. It must be either provided as another
primitive, or implemented with side-effects (to remember that $\mathit{if}$ had
solutions).  So, what does ``declarative'' mean? One might argue that
``declarative'' is really about having a compositional semantics that is simple
to understand, whether this is based in predicate logic and set-oriented or based
in outcomes and sequence-oriented.  Simple compositionality is what eases the
task of reasoning about programs, by both human programmers and meta-level
software tools.

The constructions presented in this paper are just the essential core for
a language with real-world applicability, that must include several
semantic extensions. A paramount example is arithmetic. Following
Prolog's way we must consider a (partial) denotation $^\A\denot
.:\T\times\S\imp\T$ that interprets a term under a setting as an arithmetic
expression yielding a constant term representing a number, and define
\[\denot{x\;\mtt{is}\;e}(s)=\caseif{
\denot{x\obeq y}(s)\cond{^\A\denot e(s)=y}\\
\ff\other
}\]

Another practical requirement is the ability to generate and handle
exceptions. For example, an exception is better than failure for the
semantics of a free variable task.  The formalisation requires the introduction
of a third type of final outcome, the \emph{exception} $\{t\}\in\Out^f$ with a term
$t\in\T$ conveying contextual information.  The semantic equations handling
exception in sum, product and pruning are similar to those for divergence. A
special task must be
introduced to throw an exception,\\[-13pt]
\[\denot{\mtt{throw}(t)}(\stng V\s)=\{\s(t)\}\]
and another one for catching it,\\[-13pt]
\[
\denot{\mtt{catch}(\mathit{task},\mathit{exc},\mathit{handler})}(s)=
\catch{\denot{\mathit{task}}(s)}{\mathit{exc},\mathit{handler}}
\]\\[-30pt]
\begin{alignat*}{2}
\ctch\ff &= \ff & \ctch{(\sout so)} &= \sout s{(\,\ctch o\,)}\\
\ctch\infty & = \infty & \qquad\qquad
\ctch{\{t\}} & = \caseif{
\denot h(s')\cond\denot{x\obeq t}(s)=\sout{s'\!}\ff\\
\{t\}\other
}
\end{alignat*}

Yet another unavoidable extension, in practice, is to have internal
side-effects. The required change of the semantic domain is relatively
simple, adding a persistent state alongside the setting. Lacking space here, this has to be
reported elsewhere.

\section{Conclusions and further work}

We have shown that the expressive power of Prolog can be captured with three
structural ingredients---disjunction, conjunction and \munt---plus unification,
with a simple compositional denotational semantics handling the deterministic
sequential nature of multiple solutions---equating variables to rational
trees---and final outcomes of finite failure and divergence. For the first time
the equivalent of Prolog's cut has been given compositional semantics based
solely on the state of variable bindings.  The semantics are quite naturally
extended to deal with exceptions and even side-effects, not presented here due to
space limitations.  It would be interesting to cast the semantics in a
co-algebraic account. We have also defined, but not yet reported, a more concrete
operational (step) semantics in terms of graph rewriting that nicely formalises
the so-called 4-port model introduced for Prolog debugging.

Procedures are composed from clauses with a redesigned syntax, corresponding to
abstractions of the alternative branch of either a disjunction or an \mite, the
latter being an ubiquitous programming construct that is implementable with
\munt{} (but not vice-versa). We may, therefore, express \mite{} chains across clauses, not
just within one. The formalisation of clause composition uses an extended term
syntax with variable abstraction as in the $\lambda$-calculus. This paves the way
for a more ambitious endeavour to adapt the modularity style of contextual logic
programming \cite{mo_po:89:clp} to naturally and properly handle defaults and
exceptions and higher-order procedures, a great help for building complex
applications.  An issue worth exploring is the possible combination of
sequence-based semantics with program parts having set-based semantics that can
profit from computation techniques such as tabling. Another is a classification
of procedures and call patterns according to their behaviour, and its impact on
compilation.

The ideas in the paper have been turned into a practical alternative to Prolog, easy to
program and debug, and more readily amenable to partial evaluation,
important for compile-time optimisation of clean high-level declarative code. The
language---\cube{}---has been implemented on top of a Prolog system and heavily
used to good effect in building a sophisticated large real-world application
\cite{po:03:iispp}.  It incorporates several other features such as structural
abstraction and application \cite{po:02:saalp} for higher order and functional
notation.  We currently work on its contextual modularity, for which we plan to write a modular
partial evaluator.

\bibliography{sapscs}

\end{document}

%% file: macros.tex
\input{logic}

\input{math_theorem}

\newcommand{\cube}{\textsf{Cube}}

\def\A{\mathcal A}
\def\LL{\mathcal L}
\def\OO{\mathcal O}
\def\Stng{\S}
\def\Out{\OO}
\def\Ss{\Sigma}
\def\s{\sigma}

\def\fss{\fsubset}
\def\nil{[]}
\def\applsym{\triangleleft}
\def\appl#1#2{#1\applsym#2}
\def\abstsym{\!\cdot\!}
\def\abst#1#2{\lambda#1\abstsym#2}
\def\abstt#1#2{\lambda\texttt{#1}\abstsym\texttt{#2}}

\def\fv#1{\widehat{#1}}

\def\oband{\texttt{,}}
\def\obor{\texttt{;}}
\def\obunt{\;\texttt{until}\;}
\def\obeq{\,\texttt{=}\,}
\def\munt{\textsf{until}}
\def\mite{\textsf{if-then-else}}
\def\maft{\;\textsf{after}\;}
\def\mclo{\textsf{close}\;}

\def\mtt#1{\mbox{\tt #1}}

\def\soleq#1#2{\texttt{#1}\!=\!\texttt{#2}}

\def\stng#1#2{#1\!\!:\!#2}
\def\ff{\emptyset}
\def\soutsym{{\,.\,}}
\def\sout#1#2{#1\soutsym#2}
\def\ous#1#2{#1\vdash#2}
\def\ssum{\oplus}
\def\sprod{\otimes}
\def\sprun{\oslash}

\def\unfr{\rightarrow}
\newcommand{\unfa}[3][\s]{U(#2,#3,#1)}
\newcommand{\Unf}[4][{}]{\unfa{#2}{#3}#1\unfr#4}

\newcommand{\apply}[2]{#1\bullet#2}

\newcommand{\next}[1]{\check #1}

\newcommand{\stp}{S_P}

\def\catch#1#2{#1\,\langle #2\rangle\,s}
\def\ctch#1{\catch{#1}{x,h}}


%% file: logic.tex
\input{math}

\def\cor{\vee}
\def\imp{\rightarrow}

\def\bic{\leftrightarrow}
\def\Imp{\Rightarrow}

\def\Bic{\Leftrightarrow}

\def\scor{\around\,\cor}

\def\sImp{\around\,\Imp}

\def\B{{\cal B}}               
\def\C{{\cal C}}               
\def\E{{\cal E}}               

\def\I{{\cal I}}               

\def\S{{\cal S}}               
\def\T{{\cal T}}               
\def\V{{\cal V}}               

\def\denotstart{[\![\,}
\def\denotend{\,]\!]}

\def\denot#1{\denotstart#1\denotend}

\def\encodestart{\{\!|\,}
\def\encodeend{\,|\!\}}

\def\encode#1{\encodestart#1\encodeend}



%% file: math.tex
\usepackage{latexsym}
\usepackage{amsmath}

\def\2#1{#1#1}
\def\3#1{#1#1#1}
\def\4#1{#1#1#1#1}

\def\around#1#2{#1#2#1}

\newcommand{\be}{\beta}

\def\<{\langle}
\def\>{\rangle}

\newcommand\caseif[1]{\left\{\begin{array}{ll}#1\end{array}\right.}
\newcommand\caseof[1]{\left\{\begin{array}{ll}#1\end{array}\right.}
\newcommand\cond{&\quad\!\!\mbox{if}\quad}
\newcommand\other{&\quad\!\!\mbox{otherwise}}

\def\set#1#2{\{\,#1\;|\;#2\,\}}

\def\subfin{_{\mathrm{fin}}}

\def\fsubset{\subset\subfin}

\newcommand{\dom}[1]{\mathit{dom}(#1)}
\newcommand{\map}{\rightarrow} 

\newcounter{cnlist}                        

%% file: math_theorem.tex
\newtheorem{theorem}{Theorem}

\newtheorem{definition}[theorem]{Definition}

%% file: sapscs.bbl
\begin{thebibliography}{}

\bibitem[\protect\citeauthoryear{A{\"i}t-Kaci}{A{\"i}t-Kaci}{1991}]{ai-ka:91:w%
amtr}
{\sc A{\"i}t-Kaci, H.} 1991.
\newblock {\em Warren's Abstract Machine: A Tutorial Reconstruction}.
\newblock MIT Press.

\bibitem[\protect\citeauthoryear{Andrews}{Andrews}{2003}]{And:03:wpspc}
{\sc Andrews, J.~H.} 2003.
\newblock The witness properties and the semantics of the prolog cut.
\newblock {\em Theory Pract. Log. Program.\/}~{\em 3,\/}~1, 1--59.

\bibitem[\protect\citeauthoryear{Bossi, Gabbrielli, Levi, and Martelli}{Bossi
  et~al\mbox{.}}{1994}]{bo_ga_:94:ssata}
{\sc Bossi, A.}, {\sc Gabbrielli, M.}, {\sc Levi, G.}, {\sc and} {\sc Martelli,
  M.} 1994.
\newblock The s-semantics approach: Theory and applications.
\newblock {\em The Journal of Logic Programming\/}~{\em 19/20}, 149--197.

\bibitem[\protect\citeauthoryear{Clark}{Clark}{1978}]{cl:78:nf}
{\sc Clark, K.~L.} 1978.
\newblock Negation as failure.
\newblock In {\em Logic and Databases} (New York), {H.~Gallaire} {and}
  {J.~Minker}, Eds. Plenum Press, 293--322.

\bibitem[\protect\citeauthoryear{Colmerauer}{Colmerauer}{1993}]{co:82:pif}
{\sc Colmerauer, A.} 1993.
\newblock Prolog and infinite trees.
\newblock In {\em Logic Programming}, {K.~Clark} {and} {S.-A. T\"arnlund}, Eds.
  A.P.I.C. Studies in Data Processing, vol.~16. Academic Press, 231--251.

\bibitem[\protect\citeauthoryear{de~Bruin and de~Vink}{de~Bruin and
  de~Vink}{1989}]{dBr_dVi:89:cspc}
{\sc de~Bruin, A.} {\sc and} {\sc de~Vink, E.} 1989.
\newblock Continuation semantics for {PROLOG} with cut.
\newblock In {\em {TAPSOFT} '89; Proceedings of the International Joint
  Conference on Theory and Practice of Software Development}, {J.~D{\'i}az}
  {and} {F.~Orejas}, Eds. Springer, 178--192.

\bibitem[\protect\citeauthoryear{Debray and Mishra}{Debray and
  Mishra}{1988}]{de_mi:88:dosp}
{\sc Debray, S.~K.} {\sc and} {\sc Mishra, P.} 1988.
\newblock Denotational and operational semantics for {P}rolog.
\newblock {\em Journal of Logic Programming\/}~{\em 5,\/}~1, 81--91.

\bibitem[\protect\citeauthoryear{Deransart, Ed-Dbali, and Cervoni}{Deransart
  et~al\mbox{.}}{1991}]{de_ed-db_ce:91:ps}
{\sc Deransart, P.}, {\sc Ed-Dbali, A.}, {\sc and} {\sc Cervoni, L.} 1991.
\newblock {\em Prolog: The Standard; reference manual}.
\newblock Spinger.

\bibitem[\protect\citeauthoryear{Kakas, Kowalski, and Toni}{Kakas
  et~al\mbox{.}}{1993}]{ka_ko_to:93:alp}
{\sc Kakas, A.}, {\sc Kowalski, R.}, {\sc and} {\sc Toni, F.} 1993.
\newblock Abductive logic programming.
\newblock {\em Journal of Logic and Computation\/}~{\em 2,\/}~6, 719--770.

\bibitem[\protect\citeauthoryear{Li}{Li}{1994}]{li:94:pcsp}
{\sc Li, B.~Z.} 1994.
\newblock A pi-calculus specification of {P}rolog.
\newblock In {\em European Symposium on Programming}. 379--393.

\bibitem[\protect\citeauthoryear{Monteiro and Porto}{Monteiro and
  Porto}{1989}]{mo_po:89:clp}
{\sc Monteiro, L.} {\sc and} {\sc Porto, A.} 1989.
\newblock Contextual logic programming.
\newblock In {\em Logic Programming, Proceedings of the Sixth International
  Conference}, {G.~Levi} {and} {M.~Martelli}, Eds. MIT Press, 284--299.

\bibitem[\protect\citeauthoryear{Monteiro and Porto}{Monteiro and
  Porto}{1998}]{mo_po:98:ebac}
{\sc Monteiro, L.} {\sc and} {\sc Porto, A.} 1998.
\newblock Entailment-based actions for coordination.
\newblock {\em Theoretical Computer Science\/}~{\em 192}, 259--286.

\bibitem[\protect\citeauthoryear{Naish}{Naish}{1986}]{NUProlog}
{\sc Naish, L.} July 1986.
\newblock Negation and quantifiers in {NU-Prolog}.
\newblock {\em Proceedings of the Third International Conference on Logic
  Programming\/}, 624--634.

\bibitem[\protect\citeauthoryear{Porto}{Porto}{2002}]{po:02:saalp}
{\sc Porto, A.} 2002.
\newblock Structural abstraction and application in logic programming.
\newblock In {\em Functional and Logic Programming, 6th International
  Symposium, FLOPS 2002, Proceedings}, {Z.~Hu} {and}
  {M.~Rodr{\'\i}guez-Artalejo}, Eds. Lecture Notes in Computer Science, vol.
  2441. Springer, 275--289.

\bibitem[\protect\citeauthoryear{Porto}{Porto}{2003}]{po:03:iispp}
{\sc Porto, A.} 2003.
\newblock An integrated information system powered by {P}rolog.
\newblock In {\em Practical Aspects of Declarative Languages, 5th International
  Symposium, Proceedings}, {V.~Dahl} {and} {P.~Wadler}, Eds. Lecture Notes in
  Computer Science, vol. 2562. Springer, 92--109.

\bibitem[\protect\citeauthoryear{Seres, Spivey, and Hoare}{Seres
  et~al\mbox{.}}{1999}]{se_sp_ho:99:alp}
{\sc Seres, S.}, {\sc Spivey, M.}, {\sc and} {\sc Hoare, T.} 1999.
\newblock Algebra of logic programming.
\newblock In {\em Proceedings of the 1999 international conference on Logic
  programming}. Massachusetts Institute of Technology, Cambridge, MA, USA,
  184--199.

\bibitem[\protect\citeauthoryear{Somogyi, Henderson, and Conway}{Somogyi
  et~al\mbox{.}}{1996}]{Mercury}
{\sc Somogyi, Z.}, {\sc Henderson, F.}, {\sc and} {\sc Conway, T.} 1996.
\newblock The execution algorithm of {Mercury}, an efficient purely declarative
  logic programming language.
\newblock {\em Journal of Logic Programming\/}~{\em 29,\/}~1--3, 17--64.

\bibitem[\protect\citeauthoryear{Spoto}{Spoto}{2000}]{sp:00:ogidspc}
{\sc Spoto, F.} 2000.
\newblock Operational and goal-independent denotational semantics for prolog
  with cut.
\newblock {\em Journal of Logic Programming\/}~{\em 42,\/}~1, 1--46.

\bibitem[\protect\citeauthoryear{van Emden and Kowalski}{van Emden and
  Kowalski}{1976}]{v-em_ko:76:splpl}
{\sc van Emden, M.~H.} {\sc and} {\sc Kowalski, R.~A.} 1976.
\newblock The semantics of predicate logic as a programming language.
\newblock {\em Journal of the ACM\/}~{\em 23(4)}, 733--742.

\end{thebibliography}
